\documentclass[twocolumn]{aastex63}

\usepackage{graphicx}
\usepackage{subfigure}
\usepackage{amsmath}    
\usepackage{amssymb}    
\usepackage{bm}
\usepackage{url}
\usepackage{upgreek}
\usepackage{xspace}
\usepackage[utf8]{inputenc}
\usepackage{float}

\usepackage{xcolor}
\definecolor{redl}{RGB}{255,0,0}

\shorttitle{PRS associated with FRB~20240114A}
\shortauthors{Bhusare et al.}

\begin{document}

\title{Low-frequency Probes of the Persistent Radio Sources associated with Repeating FRBs}


\correspondingauthor{Yash Bhusare}
\email{ybhusare@ncra.tifr.res.in}
\author[0000-0002-5342-163X]{Yash Bhusare}
\affil{National Centre for Radio Astrophysics, Tata Institute of Fundamental Research, Post Bag 3, Ganeshkhind, Pune - 411007, India}
\author[0000-0002-0862-6062]{Yogesh Maan}
\affil{National Centre for Radio Astrophysics, Tata Institute of Fundamental Research, Post Bag 3, Ganeshkhind, Pune - 
411007, India}
\author[0009-0002-0330-9188]{Ajay Kumar}
\affil{National Centre for Radio Astrophysics, Tata Institute of Fundamental Research, Post Bag 3, Ganeshkhind, Pune - 
411007, India}

\begin{abstract}

The discovery of Persistent Radio Sources (PRSs) associated with three repeating fast radio bursts (FRBs) has provided insight into the local environments of these FRBs. Here, we present deep radio observations of the fields surrounding three highly active repeating FRBs namely, FRB\,20220912A, FRB\,20240114A, and FRB\,20240619D using the upgraded Giant Metrewave Radio Telescope (uGMRT) at low radio frequencies. Towards FRB\,20240114A, we report the detection of compact source at 650\,MHz with a flux density of 65.6$\pm$8.1\,$\mu$Jy/beam. Our measurements of the spectral index, star formation rate of the host galaxy and recently reported constraints on the physical size strongly argue for our detected source to be a persistent radio source (PRS) associated with the FRB\,20240114A. We investigate possible origins of the PRS associated with FRB\,20240114A. Based on its brightness and age, we rule out central engines formed via accretion-induced collapse of a white dwarf, while superluminous supernovae, long gamma-ray bursts, and neutron star merger channels remain viable. An off-axis GRB afterglow could also explain the observed emission. For FRB\,20220912A, we detect radio emission that is most likely due to star formation in the host galaxy. For FRB\,20240619D, we provide upper limits on the radio emission from an associated PRS or the host galaxy. The detection of the PRS associated with FRB\,20240114A is a useful addition to the PRSs known to be associated with only three other FRBs so far, and further supports the origin of the PRS in the form of magnetoionic medium surrounding the FRB sources.
\end{abstract}

\keywords{ Fast Radio Bursts (2008) --- FRBs (2008) --- Radio Transients (1868) }

\section{Introduction} \label{sec-intro}

Since their discovery, fast radio bursts (FRBs) \citep{Lorimer_2007} have captured significant interest due to their exotic natures. FRBs are typically categorized into two types, repeaters and one-off events. Repeaters offer an excellent opportunity to study FRBs in detail, including the statistics of various burst properties, their temporal and spectral evolution and energetics \citep[e.g.,][]{zhang2023fastobservationsfrb20220912a,sand2023chimefrbstudyburstrate,Li_2019,The_CHIME_FRB_Collaboration_2023,kirsten2023connectingrepeatingnonrepeatingfast,kumar2024varyingactivityburstproperties}.
Given the millisecond durations of FRBs, their origins are likely to be compact objects. Analyzing the properties of FRBs provides insights into both the radiation mechanisms involved and the environment surrounding the compact engines that produce them. To explore and characterize the radiative environment around FRBs, deep continuum studies are crucial.

Persistent Radio Sources (PRSs) associated with FRBs are compact and characterized by continuous radio emission, unlike the transient nature of FRBs. To date, PRS associated with three FRBs have been found, FRB 20121102A \citep{Chatterjee_2017}, FRB 20190520B \citep{Niu_2022}, and FRB 20201124A \citep{bruni2024nebularoriginpersistentradio}. The non-thermal radiation from these PRSs is thought to originate from the environment surrounding the compact object driving the FRB emission. The first two detected PRSs exhibit high luminosity, of the order of $10^{29}$ erg/s/Hz \citep{Chatterjee_2017,Niu_2022}. More recently, \cite{bruni2024nebularoriginpersistentradio} reported the detection of a faint PRS associated with FRB 20201124A. The luminosities of these PRSs show a correlation with the Rotation Measure (RM) deduced from bursts, suggesting that the persistent emission is tracing a dense magneto-ionic medium. The relatively high RM observed in the corresponding FRB independently supports this scenario. 

\begin{figure*}[htbp]
    \centering
    \includegraphics[width=0.98\textwidth]{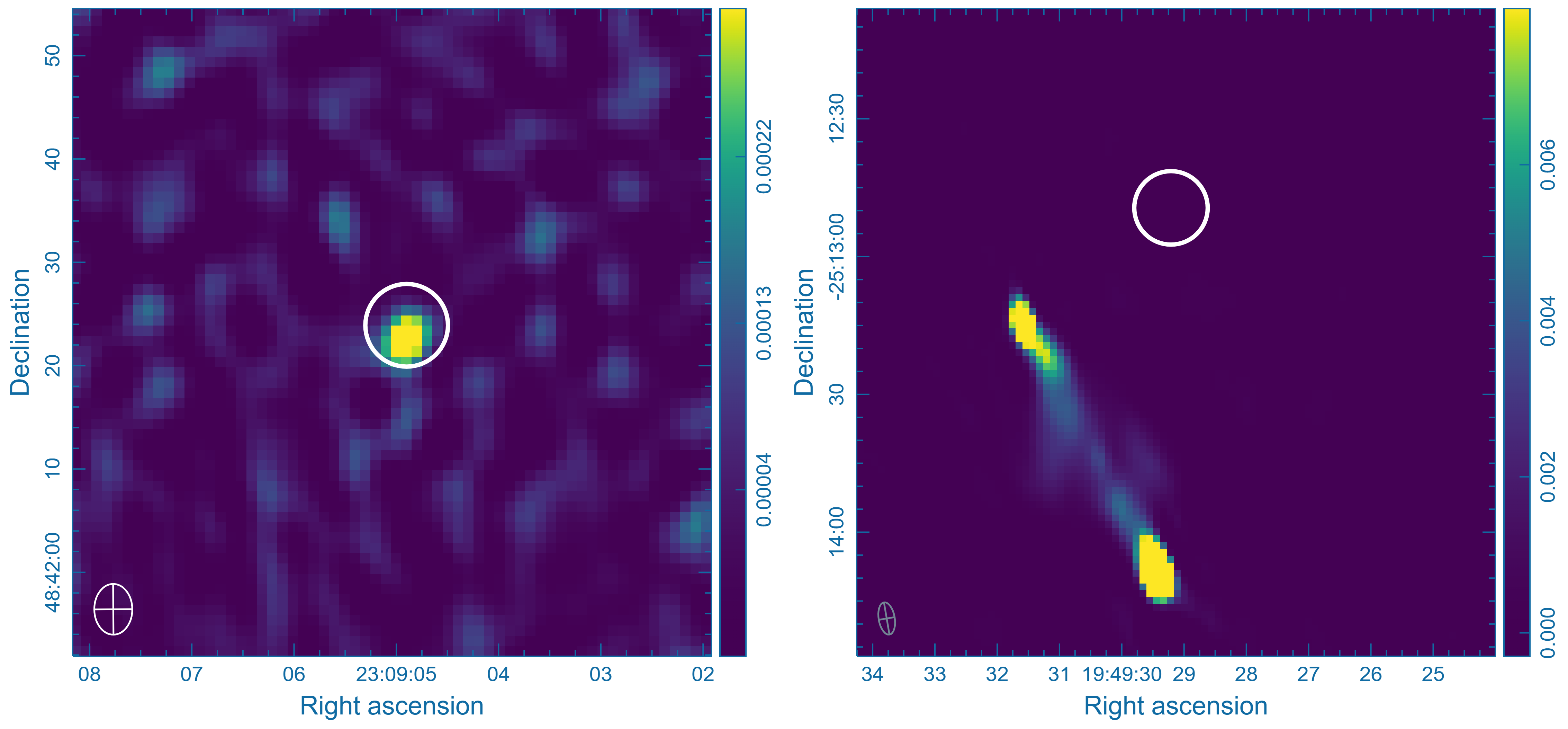}
    \caption{The uGMRT Band 4 radio images showing the fields around FRB 20220912A (left) and FRB 20240619D (right). The image of the FRB 20240619D field is limited by dynamic range due to the presence of a nearby bright AGN. Circles indicate the locations of the respective FRBs.
}
    \label{fig:R117_Image}
\end{figure*}  

Detecting and studying more PRSs could significantly advance our understanding of the origins of FRBs, their local environments, and the nature of the compact objects powering them. The faintness of the newly discovered PRS associated with FRB 20201124A raises the possibility that many more such sources exist but have remained undetected due to observational limitations. It is also possible that some FRBs could be linked to compact objects embedded in magnetized plasma, while others might be located in more diffuse media, leading to different observable properties. In this context, the study of FRB surroundings is crucial for differentiating these potential models. By probing the local environments of FRBs, we can explore the mechanisms responsible for their emission and better understand, whether the associated PRSs are a universal feature of all (repeating) FRBs or specific to certain progenitor systems.

Given the potential diversity of FRB environments, a detailed study of individual sources is essential to understand the physical mechanisms. In this work, we focus on three highly active repeaters, namely, FRB 20220912A, FRB 20240114A, and FRB 20240619D each offering a unique opportunity to investigate their surrounding environments.

FRB 20220912A, discovered in September 2022, became highly active in October 2022. CHIME/FRB detected 9 bursts from this FRB over three days \citep{2022ATel15679....1M}, indicating it is a highly active repeating source. The FRB has a dispersion measure of 219.5 pc cm$^{-3}$ and the RM is 0.6 radians m$^{-2}$. The high level of activity allowed DSA$-$110 collaboration to quickly localize the source in the host galaxy at a redshift of 0.077 \citep{Ravi_2023}. Following its discovery, it was observed by various other telescopes, including our own observations with the Giant Metrewave Radio Telescope (GMRT), resulting in the detection of over 100 bursts in November 2022 \citep{2022ATel15691....1H,2022ATel15791....1R,2022ATel15806....1B}. Due to its high activity, we have been continuing our monitoring with GMRT, and leveraging GMRTs capability to simultaneously record timeseries and visibility data (detailed analysis on burst rate and properties will be reported elsewhere).

FRB 20240114A was discovered by the CHIME/FRB Collaboration in January 2024 \citep{2024ATel16420....1S}. Based on the observations from MeerKAT \citep{10.1093/mnras/stae2013} and the European Very Long Baseline Interferometry Network (EVN) \citep{2024ATel16542....1S}, FRB 20240114A is localized (RA = 21h27m39.835s, Dec = +04d19m45.634s) in the galaxy J212739.84+041945.8 at a redshift of z=0.13 \citep{2024ATel16613....1B_redshift_mohit}. Bursts from FRB 20240114A are detected in a wide frequency range from 300 MHz to 6 GHz using various telescopes (e.g.,\citep{2024ATel16432....1O,2024ATel16599....1J,2024ATel16620....1L}), with potential hints of even chromaticity \citep{kumar2024varyingactivityburstproperties}. The reported RM of FRB 20240114A is 338 rad/m$^{2}$ \citep{10.1093/mnras/stae2013} and its DM is 527 pc cm$^{-3}$. 

Recently, \cite{2024ATel16695....1Z} reported the detection of L-band radio emission co-located with FRB~20240114A using MeerKAT, identifying a compact source with a continuum flux density of $72 \pm 14~\mu\text{Jy}$. Following this, we independently detected persistent radio emission towards FRB~20240114A using uGMRT observations at 650\,MHz, which we announced in \citet{2024ATel16820....1B}. Later, \citet{2024ATel16885....1B} detected persistent radio emission at a high frequency (5 GHz) and with higher angular resolution, confirming its compactness.

\begin{figure*}[htbp]
    \centering
    \includegraphics[width=0.98\textwidth]{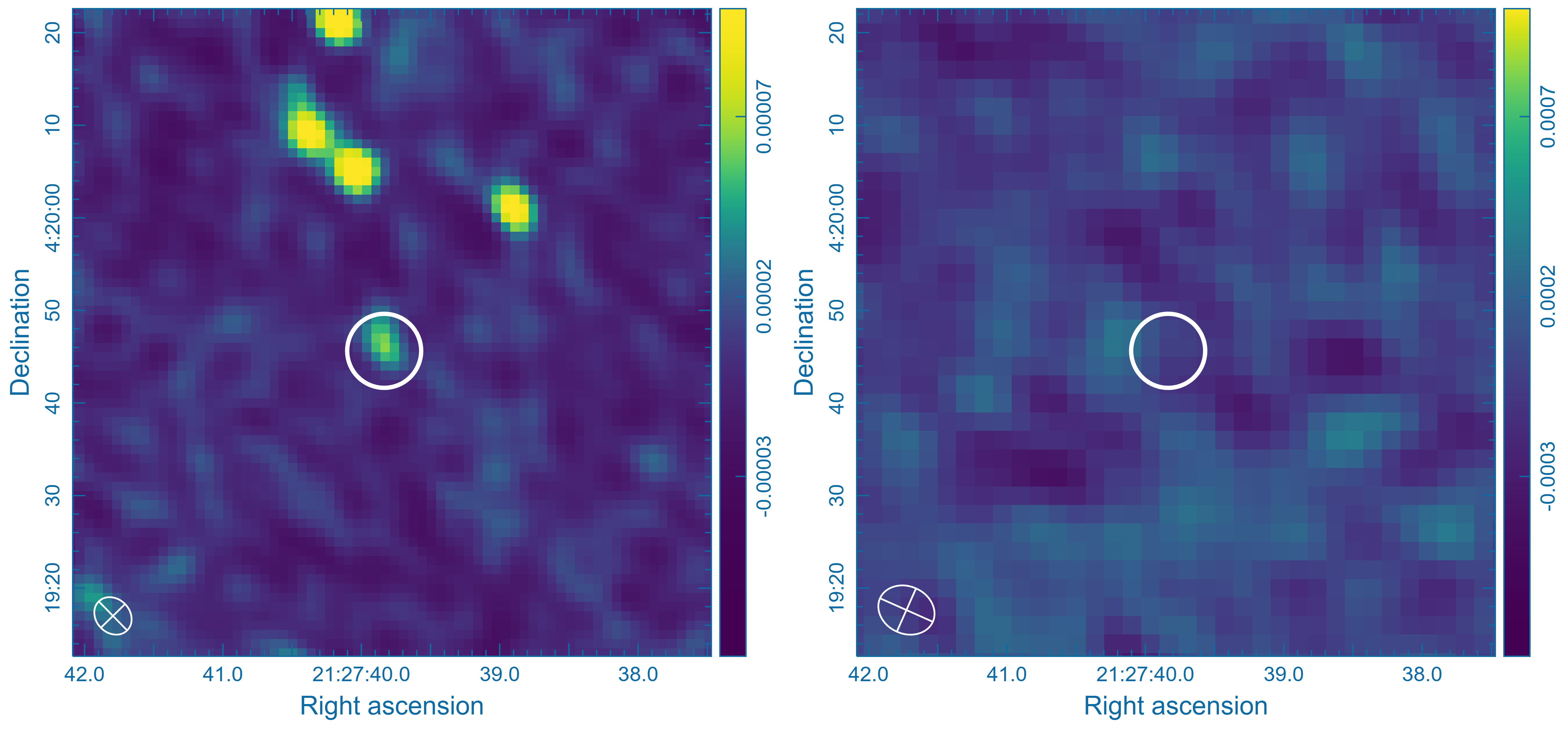}
    \caption{Left: A uGMRT Band 4 image of the field around FRB~20240114A, showing a detection of a PRS at the FRB location. Right: Band 3 image of the same field with no detection, limited by relatively large noise. White circles are centered at location of the FRB with a 4" radius. 
}
    \label{fig:R147_Image}
\end{figure*}

FRB 20240619D is a repeating source discovered by the MeerTRAP team. They detected three bursts from the source within two minutes on 19 June 2024. These bursts triggered the voltage dump from individual antennas, helping in the localization of the source \citep{2024ATel16690....1T}. The MeerTRAP team reported an RM of approximately 177 rad/m$^{2}$ and a DM of 465.48 pc cm$^{-3}$. \cite{2024ATel16745....1K} followed up with uGMRT and detected 60 bursts, confirming its activity in low frequency as well. There are two optical sources near the FRB location in the images from the DESI Legacy Survey DR10 \citep{2019AJ....157..168D}. No redshift information is available for these sources. FRB 20240619D could potentially be associated with either of them or an even fainter galaxy, as mentioned by the MeerTRAP team \citep{2024ATel16690....1T}.
 
In this paper, we present a study of the fields in the direction of these three highly active repeating FRBs using the upgraded GMRT (uGMRT). Our primary objective is to search for radio emission around these repeating FRBs that could indicate the presence of a PRS or provide clues about their surrounding environments.
In the following sections, we provide a detailed description of the observational setup, followed by the data analysis methods, and finally, we discuss our results and their implications.

\section{Observations}\label{observation}

The observations were carried out using the Band-3 (300–500 MHz) and Band-4 (550–850 MHz) of uGMRT, leveraging its capability for simultaneous timeseries and visibility recording. We have observed all the three FRBs at several epochs, and details of the specific observations used in this work are provided below.

For FRB 20220912A, we conducted observations simultaneously in Band-3 and Band-4 by using a dual sub-array mode. Approximately half the antennas were allocated to each of the frequency bands. Data were collected over eight epochs, spanning from May 1 to September 24, 2023 and 3 epochs were imaged. Throughout all epochs, 2355+498 was used as the phase calibrator. Observation was done with 4096 frequency channels and with 5.3 seconds of integration time.  

Observations of FRB 20240619D were conducted in Band-3 and Band-4 separately, on 28 August 2024. We dedicated 68 minutes of on-source time in Band-4 and 71 minutes in Band-3. The phase calibration was performed using 1833-210 and 1830-360, while 3C286 was used as the flux calibrator.   

For FRB 20240114A, we utilized a total of four Band-4 observations conducted between June 14 to August 22, 2024, with a total of 10 hours on-source time. These observations utilized 2130+050 as the phase calibrator and 3C48 as the flux calibrator. We also observed in Band-3 at 3 epochs, on 5, 8 and 24 March 2024, with the same sets of calibrators.

For both FRB 20240114A and FRB 20240619D data was recorded with integration time of 10.7 seconds and 4096 channels in band 4 and 8192 channels in band 3.

\begin{figure*}[ht]
    \centering
    \includegraphics[width=0.8\textwidth]{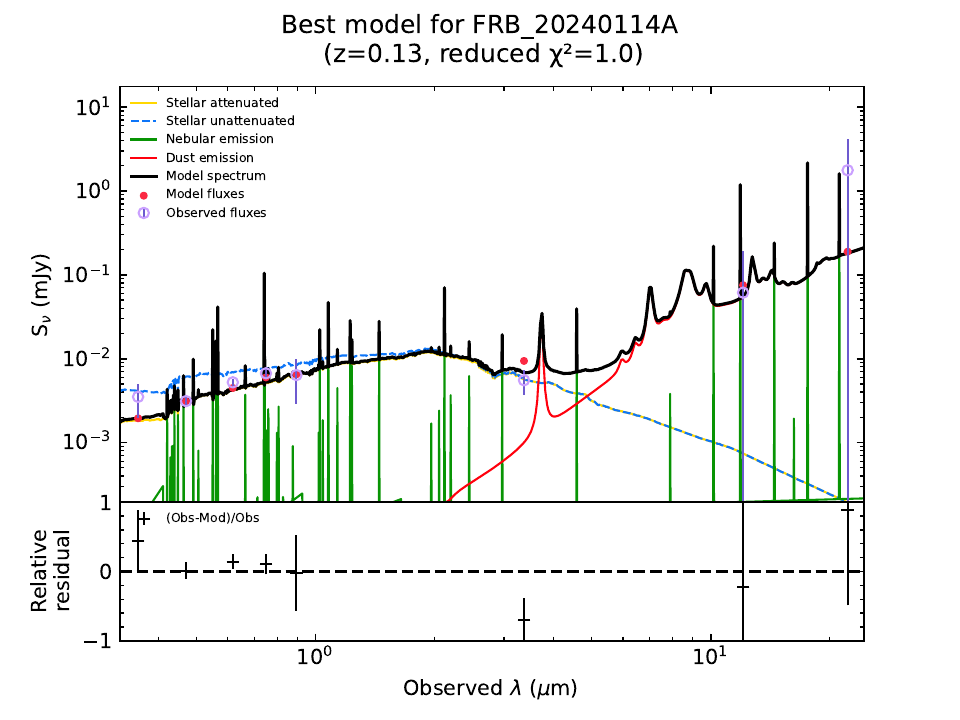}
    \caption{SED modeling of the host galaxy of FRB 20240114A using CIGALE. The observed photometric data points (purple dots) from SDSS and WISE are fitted with the model (solid black line) to estimate properties such as the SFR and the stellar mass.
}
    \label{fig:SED}
\end{figure*}

\section{Data Reduction and Analysis}\label{methods}
\subsection{Imaging}
The uGMRT observations were processed using the CAPTURE pipeline \citep{Kale_2020}, which uses the Common Astronomy Software Applications (CASA) package. The pipeline begins by initial flagging using CASA tasks like {\it tfcrop}, aimed at removing radio frequency interference (RFI). It also evaluates the statistics of each antenna, identifying and flagging any bad antennas to ensure good data quality. Following this, primary and secondary calibrations are done using the available calibrator scans. After the first round of calibration, the pipeline conducts another round of flagging, for a better RFI excision. The calibration process is then repeated on the flagged data for further refinement.

Once the calibration is complete, the pipeline processes the target field, applying additional rounds of flagging using CASA tools such as {\it rflag} and {\it tfcrop} to clean up the data further. For self-calibration, we utilize an option called {\it dosubbandselfcal}, which splits the data into different spectral windows, improving the calibration accuracy. We divided the data into 8 to 16 spectral windows during self-calibration. This is followed by multiple rounds of self-calibration, with the dynamic range of the image carefully checked after each round to ensure progressive improvement. In cases where deconvolution errors appear in the image, manual masking is applied to mitigate these issues and enhance the final image quality. 

For FRB 20220912A, our flux calibration scans were not in a usable condition, so we used phase calibrator 2355+498 also as flux calibrator. To model 2355+498 we used measurements given in VLA calibrator list. This approach assumes that flux density is not varying which may not be true. Therefore, we have accounted for an error of up to 30\% on flux measurements. The final images of FRB~20220912A and FRB~20240619D are shown in Figure~\ref{fig:R117_Image}. The white circles in both the images are drawn at the locations of the respective FRBs.

For FRB 20240114A, we conducted a deep search for a radio source by combining four epochs of data, totalling 10 hours of on-source time at Band~4. After completing all rounds of self-calibration, we merged the datasets to enhance sensitivity. Deconvolution was performed on the combined dataset using both CASA’s {\it tclean} and WSClean \citep{Offringa_2014_wsclean} separately, with both algorithms yielding similar results in terms of the residual RMS. Figure~\ref{fig:R147_Image} shows the final images from both the bands, with a continuum source detected at 650\,MHz (Band~4) and a non-detection at 400\,MHz (Band~3). Additionally, we imaged two separate parts of Band 4 from the combined data to measure the flux density at two different subbands, each having a bandwidth of 100\,MHz.

\subsection{Astrometry and Flux density estimation}
To ensure accurate astrometry, we cross-matched our dataset with the VLA FIRST survey catalogue \citep{10.1007/978-94-011-5026-2_31} for the field of FRB 20240114A. We used {\it PyBDSF } \citep{2015ascl.soft02007M} to extract all the sources detected with signal-to-noise ratio (SNR) more than 5. Following this, we conducted a cross-match of the images within a radius of $2^{\prime\prime}$ to identify corresponding sources. Then, we calculated the mean differences separately in right ascension (RA) and declination (DEC) of the matched sources, between the reference catalogue and the uGMRT image. These mean values were then used to account for any offset in the positions of the sources. For the FRB 20240114A field, the above mean astrometric offsets are $0.21^{\prime\prime}$ and $-0.09^{\prime\prime}$, with the corresponding standard deviations of $1.25^{\prime\prime}$ and $0.83^{\prime\prime}$, in RA and DEC, respectively.

We measured the flux density of any particular source using the CASA task {\it imfit}, specifying the region of the source. This task fits a Gaussian model to the source and the ratio of the peak and integrated flux density estimated from the model also informs about the compactness of the source.

\subsection{Host galaxy Properties for FRB 20240114A}

To derive the host galaxy properties for FRB 20240114A, we utilized public datasets from the Sloan Digital Sky Survey \citep[SDSS;][]{Ahumada_2020} and the Wide-field Infrared Survey Explorer (WISE) \citep{2010AJ....140.1868W}. The host galaxy was detected across all filters of SDSS, and in WISE-filters W1, W3, and W4. We used the {\it Code Investigating GALaxy Emission} \citep[CIGALE;][]{Boquien_2019_cigale} for spectral energy distribution (SED) modeling of the galaxy, enabling us to extract essential parameters such as the star formation rate (SFR) and the mass of the galaxy.

For our analysis, we used the redshift reported by \cite{2024ATel16613....1B_redshift_mohit}, $z = 0.13$. The star formation history was modeled using a double exponential function, while we utilized the \cite{10.1046/j.1365-8711.2003.06897.x} stellar population synthesis (SSP) models. Additionally, we included nebular emission and dust attenuation models based on the modified \cite{2000ApJ...533..682C} attenuation law, along with the dust emission model by \cite{2014ApJ...780..172D}.

\section{Results}\label{results}
For FRB 20220912A, we imaged data separately for three epochs with on-source times of 49, 77, and 77 minutes observed on May 1, June 2, and July 15, 2023, respectively, at Band 4. The best rms of 47 $\mu$Jy/beam was achieved from the observation on June 2, 2023. In Band 3, we got the best image from the 24\textsuperscript{th} march epoch with an rms of 350 $\mu$Jy/beam. 

We detected radio emission in the direction of FRB 20220912A at uGMRT Band 4 as shown in Figure~\ref{fig:R117_Image}. The measured flux density is $435 \pm 53 \, \mu\text{Jy/beam}$ (statistical errors only, with systematic errors potentially up to 30\%, as we did not have a reliable flux calibrator). No emission was detected at Band 3 of the uGMRT above a $5 \sigma$ level of 1.750 mJy/beam. To estimate the spectral index, we used the publicly available dataset of EVLA from proposal code 22B-307 at C-band. This data was already pre-processed using the VLA pipeline and we directly used the final image to estimate the flux density at C-band to be $89 \pm 3.6 \, \mu\text{Jy/beam}$, with an rms of $3 \mu\text{Jy/beam}$. Using this measurement at C-band and our measurement at Band 4 of uGMRT, combined with the uGMRT measurement of  $240 \pm 36 \,  \mu\text{Jy/beam}$  at 1.26\,GHz \citep{2024A&A...690A.219P} we estimate the spectral index ($\alpha$) to be $-0.73$. Previous Very Long Baseline Interferometry (VLBI) observations by \cite{hewitt2023milliarcsecondlocalisationhyperactiverepeating} did not detect any compact emission from this source above a $5\sigma$ level of  $80\,\mu\text{Jy}$, leading us to conclude that majority of the radio emission we have detected at Band 4 is likely associated with star formation activity in the host galaxy. 

The image towards FRB 20240619D has an rms of 83.4 $\mu$Jy/beam. The FRB position is very close to an Active Galactic Nucleolus (AGN) as shown in figure \ref{fig:R117_Image}. Due to this nearby bright source, our radio image is limited in dynamic range and we could not probe deep to look for any PRS. Nevertheless, we put a 5$\sigma$ upper limit of 417 $\mu$Jy/beam on a PRS associated with this FRB at 650\,MHz. At Band 3, deconvolution errors due to a bright source located between the Full Width at Half Maxima (FWHM) and the first minima of the beam did not allow us to obtain a reasonable quality image.  

For FRB 20240114A, our SED modeling of the host galaxy suggests a star formation rate (SFR) of $0.30 \pm 0.02 \, \text{M}_\odot \, \text{yr}^{-1}$. Figure \ref{fig:SED} shows the best-fit SED of the host galaxy, modeled using CIGALE. The observed photometric data points (purple dots) are fitted with the SED model. The resulting model spectrum is in black. The lower panel shows relative residuals between the observed and modeled fluxes, indicating the goodness of the fit.

Using our radio observations, we report an $8 \sigma $ detection of continuum emission at 650\,MHz, potentially from PRS associated with FRB 20240114A. Figure \ref{fig:R147_Image} shows the images at band 3 and band 4. We combined observations from 4 epochs between 14 June and 22 August 2024, amounting to a total on-source time of 10 hours. We detected a point source with a peak flux density of $65.6 \pm 8.1\, \mu\text{Jy/beam}$. We assume flux density uncertainty of 10\% \citep{2017ApJ...846..111C} in quadrature with fitting uncertainty for calculating the final error on the flux density. The location of this point source coincides with the location of the FRB. We detected no emission at the FRB position in the Band 3 image. Our best Band 3 image was from the 24 March epoch with an on-source time of 78 minutes, suggests a $5 \sigma$ upper limit of $ 600  \mu\text{Jy/beam}$. The RMS noise in the Band 3 image is limited by the deconvolution errors caused by a bright source ($325 \text{ mJy}$) located between the FWHM and the first null of the uGMRT primary beam.

The associated source with FRB 20240114A is detected in our uGMRT Band 4 image with a best-fit position of 
RA = 21h27m39.84s $\pm$ 0.14 arcsec, 
Dec = +04d19m46.68s $\pm$ 0.29 arcsec, obtained by fitting a gaussian using {\it imfit} task in CASA. The synthesized beam size is $4.7^{\prime\prime} \times 3.8^{\prime\prime}$.  We estimate the systematic position uncertainties in the uGMRT image as the standard deviations of the scatter in RA and Dec offsets from the FIRST cross-match to be $1.25^{\prime\prime}$ and $0.83^{\prime\prime}$, respectively, as mentioned in Section~\ref{methods}. The EVN localization of the FRB is
RA = 21h27m39.835s, 
Dec = +04d19m45.634s with uncertainty of 0.200 arcsec \citep{2024ATel16542....1S, 2025arXiv250611915B}. 

The sky positions of the potential PRS and the FRB are in excellent agreement with each other in RA, and consistent with each other within $1.2\sigma$ in Dec. This provides evidence supporting the physical association of the potential PRS with FRB~20240114A.

\begin{figure*}[ht]
    \centering
    \includegraphics[width=0.99\textwidth]{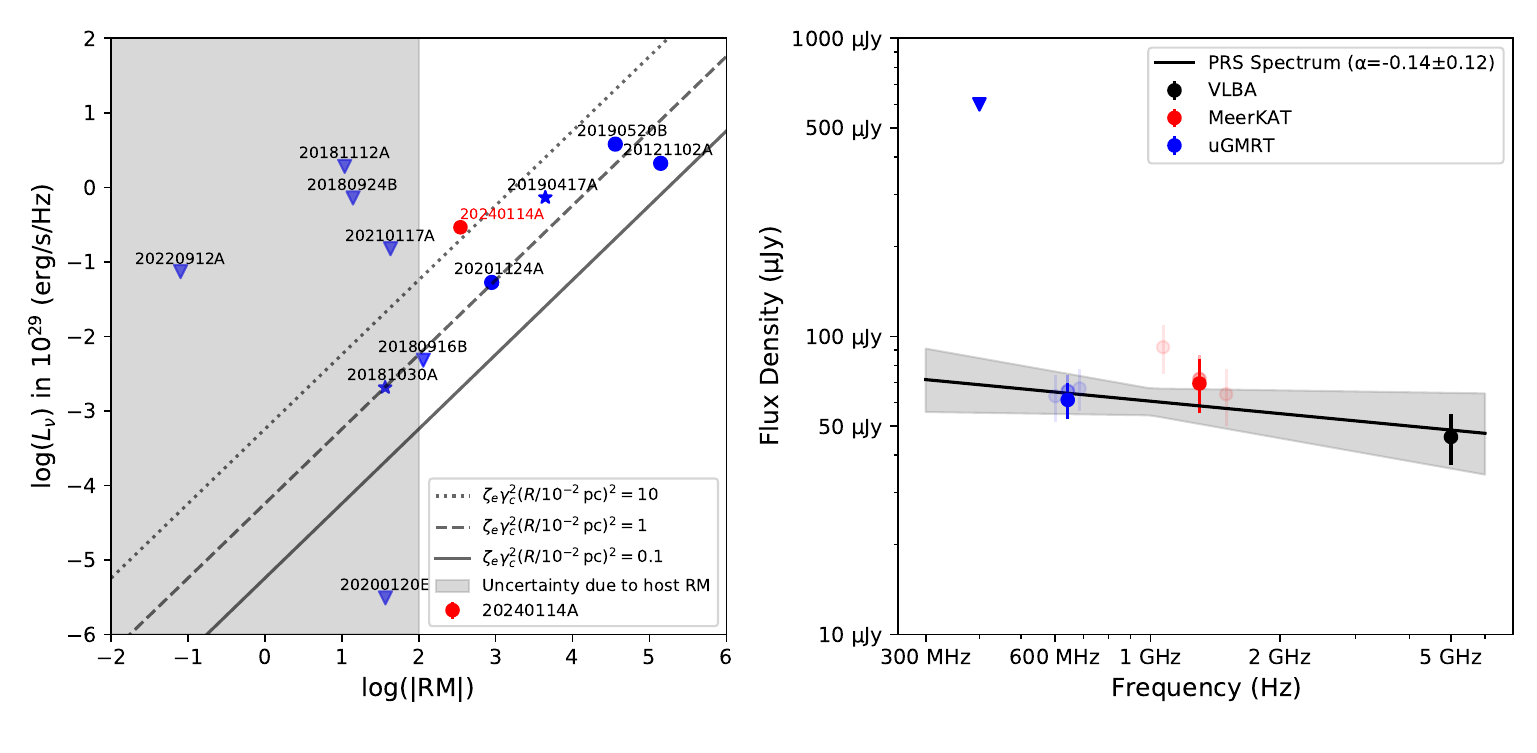}
    \caption{Left: A logarithmic plot of the luminosity versus absolute RM adapted from \cite{bruni2024nebularoriginpersistentradio}, with the addition of FRB~20240114A. The PRS luminosity at 650\,MHz is plotted after subtracting the expected contribution from the host galaxy. At low RM values (the shaded region; nominally for RM$<100$\,$rad\,m^{-2}$), it is difficult to estimate the host contribution unambiguously. We have also included potential PRS candidates associated with FRB~20190417A and FRB~20181030A \citep{2024ApJ...976..199I} denoted by stars.
    Right: The radio spectrum of the candidate PRS associated with FRB~20240114A, using measurements from this work at 650\,MHz, from \cite{2024ATel16695....1Z} at 1.3\,GHz and from \cite{bruni2024discoveryprsassociatedfrb} at 5\,GHz. The spectral index is derived from the flux density measurements corrected for the host galaxy emission (dark, filled circles). The fainter points (almost overlapping with the dark points) represent observed values without this correction, while more faded points show the sub-banded measurements, wherever available.
}
    \label{fig:prs_}
\end{figure*}

\section{Discussion}\label{discussions}

\subsection{Radio emission co-located with FRB~20240114A: PRS or star formation?} 
As a part of our analysis, we estimated the star formation rate (SFR) for the host galaxy of FRB 20240114A using {\it CIGALE} SED fitting and derived an SFR of $0.30\pm 0.02 \, M_{\odot} \text{yr}^{-1}$. However, we note that the SFR derived from Stellar Population Synthesis (SPS) using photometric measurements could have limited accuracy due to several factors, e.g., the broadband nature of the filters and dependence on the models used. On the other hand, spectroscopic measurements of the H$\alpha$ provide a more reliable SFR estimate. Based on spectroscopic data, SFR of the host galaxy is estimated to be $0.06 \pm 0.01 M_{\odot} \text{yr}^{-1}$ \citep{2025ApJ...980L..24C, 2025arXiv250611915B} indicating that our measurement using the photometric data could indeed be an overestimate. Thus, we have used the SFR of $0.06\,M_{\odot} \text{yr}^{-1}$ in any further related discussion or analysis in the rest of this paper.
\par
Using the relation by \cite{10.1111/j.1365-2966.2009.15073.x} with the above SFR estimate and an assumed spectral index of $-0.8$, we obtain expected radio flux density of $4.3 \pm 0.7\, \mu\text{Jy}$ at $650 \, \text{MHz}$ and $2.3 \pm 0.4\, \mu\text{Jy}$ at $1.4 \, \text{GHz}$. The quoted uncertainties include contribution primarily from the corresponding uncertainty on the SFR measurement. The uncertainty on the redshift $(z = 0.1306 \pm 0.0002)$ contributes less than $1\%$ to the inferred flux density and is therefore negligible. We have assumed a spectral index of $-0.8$. While the exact spectral index might vary slightly among galaxies, we note that even a change of 0.1 in the spectral index would affect the flux density estimates by less than $10\%$, which won't affect our results discussed below.
\par
The detected radio emission at 650\,MHz is much higher than the above expected from star formation alone --- the emission expected from star formation alone at 650 MHz is only about 6.5\% of the observed value. This excess radio flux density suggests the presence of a potential PRS. We subtract the expected galactic emission due to the star formation ($4.3 \pm 0.7 \, \mu\text{Jy}$) from the measured flux density ($65.6 \pm 8.1 \, \mu\text{Jy}$), resulting in the flux density of the potential PRS to be $61.3 \pm 8.1 \, \mu\text{Jy}$.

\par
Some of the earlier searches for any co-located continuum radio emission from FRB 20240114A \citep[e.g.,][]{kumar2024varyingactivityburstproperties, panda2024lowfrequencywidebandstudyactive} did not reach the required sensitivity to detect the faint emission. However, \cite{2024ATel16695....1Z} detected co-located radio continuum at L-band using MeerKAT with a flux density of $72 \pm 14  \mu\text{Jy}$. Combining our measurement at 650\,MHz and the above L-band measurement indicates towards a flat spectrum. Overall, the flat spectrum and the huge excess of radio emission when compared with that expected from the host star formation, makes a strong case for the detected emission to be from a PRS. The more recent detection of a compact source at 5\,GHz using VLBA, with a flux density of $46 \pm 9  \mu\text{Jy}$ and the physical size constrained to $\lesssim 4$\,pc \citep{bruni2024discoveryprsassociatedfrb}, confirms the radio emission to be originated from a PRS associated with FRB 20240114A.

\subsection{Origin of the PRS associated with FRB~20240114A}

The 5\,GHz detection complements our lower-frequency uGMRT observations. Our uGMRT Band 3 image yielded a non-detection with a $5\sigma$ upper limit of $600 \,\mu\text{Jy}$. In Figure~\ref{fig:prs_}, we show the flux density measurements available from \citet{2024ATel16695....1Z} and \citet{bruni2024discoveryprsassociatedfrb}, and the measurements and upper limit from this work. Figure \ref{fig:prs_} also shows the flat spectral nature of the source with a fitted spectral index of $\alpha = -0.14 \pm 0.12$, aligning with the characteristics of the other known PRSs. The sub-banded data from uGMRT and MeerKAT might suggest a potential spectral turnover around 1 GHz, however, more precise measurements and ideally at more frequencies would be needed to firmly confirm such a turnover. Nevertheless, if such a turnover is present, it could be due to the synchrotron self-absorption (SSA) or free-free absorption (FFA) from the surrounding environment of the FRB. Identifying the cause, SSA or FFA, would help in further determining the characteristics of the FRB environment. SSA would indicate a dense, magnetized region, possibly around a young neutron star or magnetar. FFA, on the other hand, would point towards a dense ionized medium, like a supernova remnant or HII region, which might be along the line of sight.  In any case, a sensitive and high spatial resolution study of the host galaxy would help in drawing any firm conclusions regarding the spectrum of the candidate PRS.

Recent broadband observations by \citet{zhang2025} reveal a spectral peak around $\sim1$~GHz, consistent with SSA. With the assumption of equipartition, they infer that the magnetic field strength in the PRS is higher than the line-of-sight magnetic field component constrained using the FRB Faraday rotation, indicating a highly magnetized central engine. They also estimate a compact source radius of $R \sim 0.03$~pc and report a strong flare, with the flux density varying by nearly 100\%, suggesting an intrinsically dynamic environment. Our investigations to probe variability at low radio frequencies (650\,MHz), using our own observations as well as those available publicly, remained inconclusive, primarily due to the low-significance detections at the individual epochs (see more details in Appendix~\ref{app1}).

\subsubsection{Magnetised plasma environment}

\begin{figure}[htbp]
  \centering
  \includegraphics[width=\columnwidth]{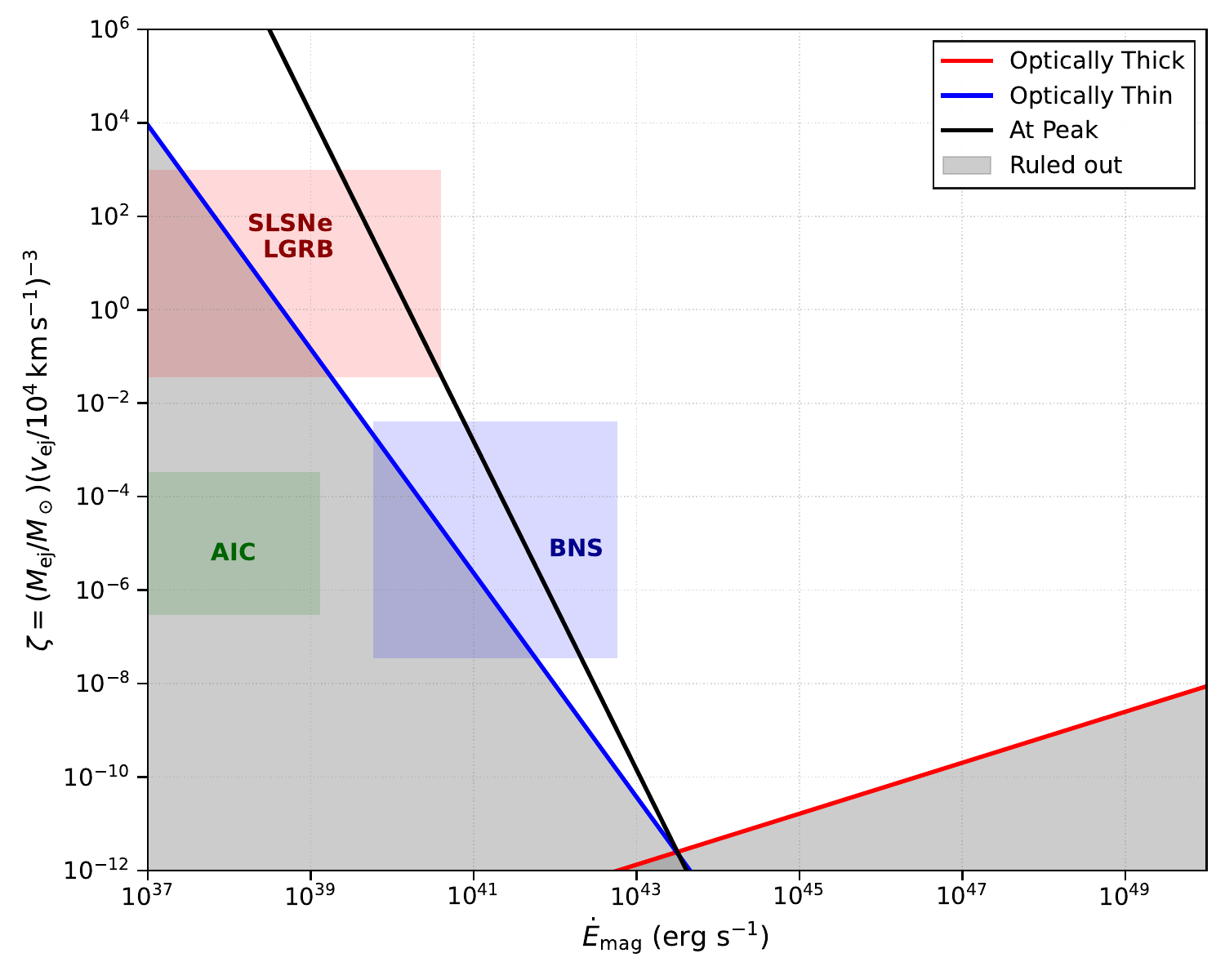}
  \caption{
  The allowed regions in the energy injection rate and the ejecta-density ($\dot{E}_{\rm mag} - \zeta$) parameter space, for an assumed minimum nebula age of 28\,days. 
  The shaded rectangular bands indicate the parameter ranges for different formation channels (SLSNe, LGRBs, AIC, and BNS), as shown in Figure~4 of \citet{2019ApJ...886..110M}. 
  The gray region is the excluded parameter space from radio detection at 650\,MHz by considering a minimum age of 28 days.}
  \label{fig:Edot_zeta}
\end{figure}

A magnetised plasma environment surrounding a FRB source can give rise to continuum emission. Such an environment is consistent with a supernova remnant or a pulsar/magnetar wind nebula \citep{2022ApJ...928L..16Y}.

Using a luminosity distance of $630.72 \text{ Mpc}$ \citep[assuming Planck18 cosmology;][]{2022ApJ...935..167A,2020}, we estimate the spectral luminosity of the PRS to be $2.9(4) \times 10^{28}$\,erg/s/Hz. This estimate indicates that the PRS associated with FRB~20240114A is about an order of magnitude fainter compared to those associated with FRB 20121102A and FRB 20190520B. We also analyzed the PRS candidate in the context of the luminosity (${L}_{\nu }$) and rotation measure (RM) relation described by \cite{2022ApJ...928L..16Y}:

\begin{equation}
    {L}_{\nu }\propto ({\zeta }_{{\rm{e}}}{\gamma }_{{\rm{c}}}^{2}{R}^{2})\times | {\rm{RM}}|
    \label{equ:lumvsrmrel}
\end{equation}

where, ${\zeta }_{\rm{e}}$ represents the fraction of electrons emitting synchrotron radiation in the observed frequency band among all electrons, ${\gamma }_{\rm{c}}$ denotes the Lorentz factor and R is the radius of the plasma contributing to emission of the PRS as well as the RM.

\begin{figure*}[ht]
    \centering
    \includegraphics[width=\textwidth]{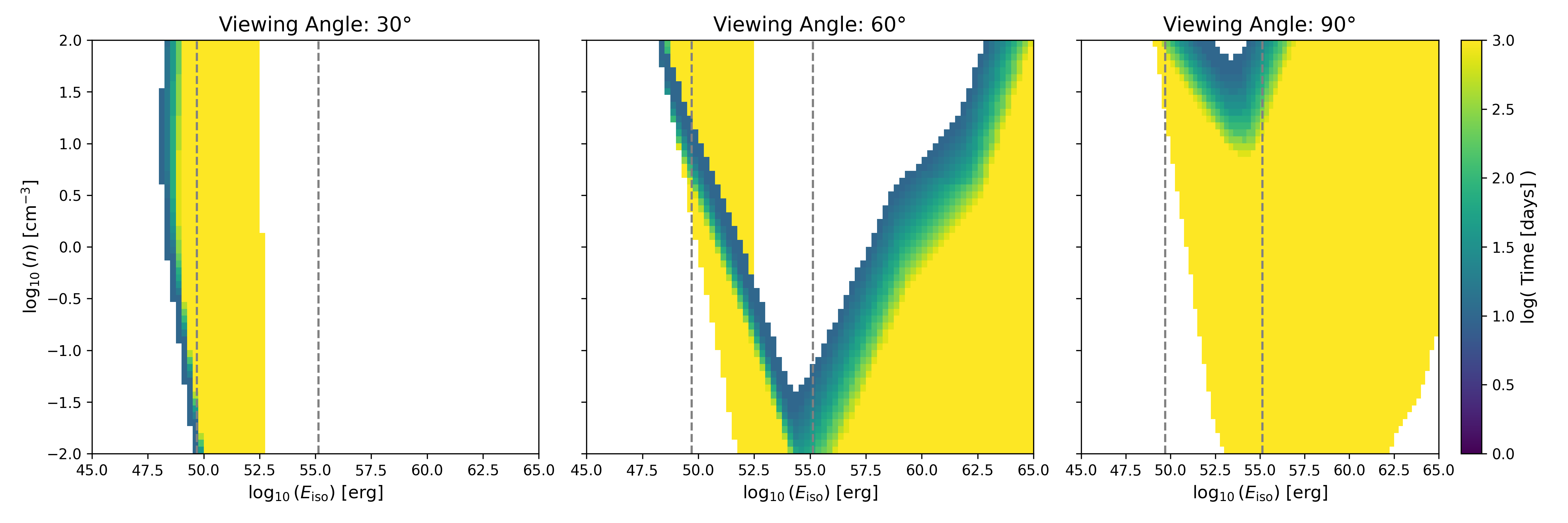}
    \caption{
    The allowed parameter space for the off-axis GRB afterglow models consistent with the measured 650\,MHz flux density of the PRS associated with FRB~20240114A. Shaded regions show the combinations of circumburst density, $n$, and isotropic energy, $E_{iso}$, that reproduce our measurement within $3\sigma$ uncertainties, for viewing angles of $\theta_{obs}=30^\circ, 60^\circ,$ and $90^\circ$. The colors indicate the time at which the model light curve intersects the observed flux density. The dotted vertical lines show the typical range of observed $E_{\mathrm{iso}}$.}
    
    
    \label{fig:Elimits}
\end{figure*} 
The other known PRSs are shown to be constrained with $0.1<{\zeta }_{{\rm{e}}}{\gamma }_{{\rm{c}}}^{2}{(R/(1{0}^{-2}{\rm{pc}}))}^{2}<10$ \citep{bruni2024nebularoriginpersistentradio}. After subtracting the estimated star formation contribution, the PRS candidate for FRB~20240114A is at the upper edge of the above parameter range, as shown in Figure~\ref{fig:prs_} \citep[assuming the observed RM of $340.5\pm0.5$\,$rad\,m^{-2}$ to be entierely due to the host;][]{10.1093/mnras/stae2013}. Therefore, the PRS associated with FRB~20240114A also follows the constraints followed by the other known PRSs, indicating that the detected emission likely originates from the magneto-ionic region surrounding the FRB, potentially in the form of a magnetar wind nebula. 


Next, we consider the implications of the observed emission within the magnetar wind nebula scenario. Following the approach taken by \citet{2024ApJ...976..199I}, we explore the allowed formation channels for the central engine using the observed flux density and age estimates of the PRS. \citet{2019ApJ...886..110M} have derived analytical solutions for the nebular emission with a simple assumption of temporally constant energy and mass injection rate from a central engine. Equations 19, 21, and 23 from \citet{2019ApJ...886..110M} express the radio luminosity in terms of the density of the surrounding ejecta and the energy injection rate from the central magnetar in the optically thick, thin and peak regimes, respectively. 
The analytical equations constrain the ejecta density, parameterized as $\zeta = (M_{\rm ej}/M_{\odot}) \times (v_{\rm ej}/10^{4}\,{\rm kms^{-1}})^{-3}$, for a given energy injection rate, $\dot{E}_{\rm mag}$. Here, $M_{\rm ej}$ is the ejecta mass and $v_{\rm ej}$ is the ejecta velocity. Assuming the age of the nebula to be longer than $28$ days, corresponding to the earliest available MeerKAT observation after the FRB discovery \citep{2024ATel16695....1Z}, we explore the allowed $\zeta - \dot{E}_{\rm mag}$ parameter space in Figure~\ref{fig:Edot_zeta}. The allowed parameter space suggests that the PRS is consistent with magnetars formed via the superluminous supernovae (SLSNe), long gamma-ray bursts (LGRBs), and binary neutron star (BNS) merger channels. Figure \ref{fig:Edot_zeta} also shows that the accretion-induced collapse (AIC) of a white dwarf as a formation channel is ruled out for the minimum assumed age of 28 days. We note that the AIC channels would be ruled out even if the age were to be as short as just a day!

\subsubsection{Gamma-ray burst afterglow}
Recent works (e.g., \citealt{2024ApJ...976..199I}) have explored the possibility that PRSs associated with FRBs might arise from GRB afterglows viewed off-axis. To examine whether the PRS associated with FRB~20240114A could be explained within such a framework, 
we simulated GRB afterglows using \texttt{afterglowpy} \citep{2020ApJ...896..166R}, adopting a Gaussian structured jet with a core angle $\theta_{\rm core}=0.17~{\rm rad}$, truncation angle $\theta_{\rm wing}=5\,\theta_{\rm core}$, the microphysical parameters $\epsilon_e=0.005$, $\epsilon_B=0.01$, and the electron power-law index $p=2.6$ (motivated by \citealt{2024ApJ...976..199I}). We explored wide parameter ranges of isotropic energies ($E_{iso} = 10^{45} - 10^{65}$ erg) and circumburst densities ($n = 10^{-2} - 10^{2}~{cm^{-3}}$), for observer viewing angles $\theta_{obs}=30^\circ, 60^\circ,$ and $90^\circ$. For each parameter combination, we computed afterglow light curves for ages ranging from $0$ to $1000$ days, and compared them with our measured 650\,MHz flux density, requiring consistency within $3\sigma$ uncertainties.

The resulting allowed regions of the parameter space are shown in Figure~\ref{fig:Elimits}. The colors indicate the time (in days) at which the simulated light curve is consistent with our measured flux density. For regions of parameter space where multiple ages are consistent with the measured flux density, we show the maximum age in the color map. For $\theta_{ obs}=30^\circ$, the allowed region is restricted to relatively lower isotropic energies, whereas for the larger viewing angles (60° and 90°), a broader range of $E_{iso}$ and $n$ combinations are permitted. The dotted vertical lines in Figure~\ref{fig:Elimits} mark the typical observed range of $E_{iso}$ values from GRBs \citep{2014ApJS..211...13V, 2014ApJS..211...12G, 2020ApJ...893...46V,2016ApJS..223...28N}. We also repeated this analysis by combining our measurement at 650\,MHz with the multi-frequency measurements (at 1.5, 3, and 5 GHz) reported by \citet{zhang2025}. As these measurements were not strictly simultaneous, we adopted a conservative time tolerance of $\pm5$ days when checking for consistency between the simulated light curves and the measured flux densities at the corresponding frequencies. With this extended data set, we found that the allowed parameter combinations were consistent with the typical observed range of $E_{iso}$ for GRBs for $\theta_{\rm obs} = 30^\circ, 60^\circ $ and $90^\circ$ (not shown in Figure~\ref{fig:Elimits}, but assessed separately).

Importantly, most of the allowed parameter space (in terms of $E_{iso}$) overlaps significantly with that for the known GRB afterglow population, indicating that an off-axis GRB afterglow could plausibly explain the observed PRS emission.

\subsection{FRB 20220912A and FRB 20240619D}
We detected radio emission towards FRB 20220912A using uGMRT observations at 650\,MHz. As the previous VLBI observations \citep{hewitt2023milliarcsecondlocalisationhyperactiverepeating} did not detect any compact radio emission associated with this FRB above a $5\sigma$ level of $80\,\mu$Jy, our uGMRT detection indicates that majority of the detected emission is likely due to star formation within the host galaxy. The SFR for this galaxy, as estimated by \cite{Ravi_2023}, is greater than 0.1 M$_{\odot}$ yr$^{-1}$. Moreover, the dust-obscured nature of the host galaxy, supported by its position on the WISE color-color plot, also explains the limited optical detections. 

The PRS detected by \cite{bruni2024nebularoriginpersistentradio} has an inverted spectrum and was found in a star-forming galaxy. Assuming a similar scenario for FRB~20220912A, the star formation emission component could have a steep spectrum, while the PRS contribution might have an inverted or flat spectrum. This would result in the star formation emission dominating at low frequencies, hindering the ability to detect any PRS emission. 
Our detected radio emission with a spectral index of -0.73 could then indeed correspond to the star formation contribution. Observations at higher frequencies and with higher spatial resolutions would be useful to detect any underlying PRS associated with FRB 20220912A with a flat or inverted spectrum. 

For FRB 20240619D, the resulting radio image has an RMS noise level of  83.4 $\mu$Jy/beam. Unfortunately, a nearby active galactic nucleus (AGN), as shown in Figure \ref{fig:R117_Image}, introduces dynamic range limitations, preventing deeper imaging of the field from searching for any possible PRS. This bright neighbouring source hampers our ability to achieve the desired sensitivity, limiting the detection of any faint PRS. Nevertheless, we put a $5\sigma$ upper limit of 417 $\mu$Jy/beam on a PRS associated with this FRB at 650 \,MHz. 

\section{Summary}\label{summary}
We have presented deep low-frequency observations of three highly active repeating FRBS, FRB 20220912A, FRB 20240114A, and FRB 20240619D, using the uGMRT. For FRB~20240114A, we report an $8\sigma$ detection of a PRS at 650\,MHz. The flat radio spectrum, low SFR of the host galaxy and the recently reported physical compactness at 5\,GHz corroborate well for the source to be a PRS associated with FRB~20240114A. The measured flux density and a flat spectrum are also consistent with the other known PRSs. The luminosity deduced from our measurements and the RM reported elsewhere, are consistent with the radio emission to be produced by a magnetoionic medium around the FRB source. We analyzed our results in the context of different progenitor models. For a magnetised plasma environment with a minimum age of 28 days, we examined the allowed regions in the magnetic energy injection rate and the ejecta density parameter space, and ruled out AIC as a possible formation channel for the central engine of FRB~20240114A, while SLSNe, LGRBs, and BNS channels remain viable. We also explored the GRB afterglow scenario to assess whether the PRS emission could arise from an off-axis GRB, and we find that, for typical GRB parameters, this scenario is indeed possible. For FRB 20220912A, we detected radio emission at 650 MHz, which is consistent with star formation in its host galaxy, as supported by its spectrum as well as the lack of observable compact emission in the previous VLBI studies. In the case of FRB 20240619D, we set upper limit on emission from an associated PRS. Future high-resolution VLBI and multi-frequency observations will be the key to further characterize the properties of the PRS associated with FRB~20240114A and to potentially uncover PRSs associated with the other two FRBs.
\acknowledgments
We would like to thank Shriharsh Tendulkar and Visweshwar Ram Marthi for the insightful discussion on FRB 20220912A at the initial stages. We sincerely acknowledge Ujjwal Panda, PI of the programs DDTC333, DDTC351, and DDTC353, the imaging data from which are discussed in the appendix. YB would like to thank Ramananda Santra and Ruta Kale for valuable discussions on interferometric data analysis, and Pralay Biswas, Rashi Jain and Yogesh Wadadekar for their insights on SED fitting. We also thank the anonymous referee whose comments helped in improving the manuscript. YM acknowledges support from the Department of Science and Technology via the Science and Engineering Research Board Startup Research Grant (SRG/2023/002657). We would like to thank the Centre Director and the observatory for the prompt time allocation and scheduling of our observations. GMRT is run by the National Centre for Radio Astrophysics of the Tata Institute of Fundamental Research. We acknowledge the Department of Atomic Energy for funding support, under project 12$-$R\&D$-$TFR$-$5.02$-$0700. This publication makes use of data products from the Wide-field Infrared Survey Explorer, which is a joint project of the University of California, Los Angeles, and the Jet Propulsion Laboratory/California Institute of Technology, funded by the National Aeronautics and Space Administration. The National Radio Astronomy Observatory is a facility of the National Science Foundation operated under cooperative agreement by Associated Universities, Inc.
Funding for the Sloan Digital Sky Survey V has been provided by the Alfred P. Sloan Foundation, the Heising-Simons Foundation, the National Science Foundation, and the Participating Institutions. SDSS acknowledges support and resources from the Center for High-Performance Computing at the University of Utah. SDSS telescopes are located at Apache Point Observatory, funded by the Astrophysical Research Consortium and operated by New Mexico State University, and at Las Campanas Observatory, operated by the Carnegie Institution for Science. The SDSS web site is \url{www.sdss.org}. SDSS is managed by the Astrophysical Research Consortium for the Participating Institutions of the SDSS Collaboration, including Caltech, The Carnegie Institution for Science, Chilean National Time Allocation Committee (CNTAC) ratified researchers, The Flatiron Institute, the Gotham Participation Group, Harvard University, Heidelberg University, The Johns Hopkins University, L'Ecole polytechnique f\'{e}d\'{e}rale de Lausanne (EPFL), Leibniz-Institut f\"{u}r Astrophysik Potsdam (AIP), Max-Planck-Institut f\"{u}r Astronomie (MPIA Heidelberg), Max-Planck-Institut f\"{u}r Extraterrestrische Physik (MPE), Nanjing University, National Astronomical Observatories of China (NAOC), New Mexico State University, The Ohio State University, Pennsylvania State University, Smithsonian Astrophysical Observatory, Space Telescope Science Institute (STScI), the Stellar Astrophysics Participation Group, Universidad Nacional Aut\'{o}noma de M\'{e}xico, University of Arizona, University of Colorado Boulder, University of Illinois at Urbana-Champaign, University of Toronto, University of Utah, University of Virginia, Yale University, and Yunnan University.

\bibliography{ref}{}
\bibliographystyle{aasjournal}

\appendix
\section{Variability of the PRS associated with FRB~20240114A at 650\,MHz}\label{app1}
To investigate any variability of PRS associated with FRB~20240114A at 650~MHz, we analysed all available uGMRT observations of FRB~20240114A, including both, our own data and those discussed in \citep{panda2024lowfrequencywidebandstudyactive} and now publicly available. In total, we analysed seven epochs spanning seven months (see Table~\ref{tab:prs_epochs}). We followed a similar approach for the imaging analysis of all the epochs as mentioned in Section \ref{methods}. Different epochs had varying on-source times, and the detection SNR at individual epochs was low, with a maximum achieved SNR of only 5.5, as shown in Table \ref{tab:prs_epochs}. Note that the flux density quoted in Table~\ref{tab:prs_epochs} is the peak flux density at the location of the PRS and not from the {\it imfit} task of CASA, as the SNR is quite low. The quoted uncertainty is only the RMS uncertainty, and does not include any systematic uncertainty. 

We performed a chi-square test to assess whether the measured flux densities across the seven epochs are consistent with a constant value. Using the weighted mean flux density as the reference, we obtained a p-value of 0.2. Given the relatively low SNR at most of the epochs, this result does not allow us to make a strong statement about variability: the observed scatter is consistent with statistical fluctuations, but we cannot rule out intrinsic variability within the large uncertainties, some of which might be dominated by systematics. Future, deeper, and higher-cadence monitoring will be crucial to robustly probe whether the variability seen at higher frequencies extends to the low frequencies.

\begin{table}[h]
\centering
\caption{uGMRT Band~4 peak flux density measurements at the PRS position.}
\label{tab:prs_epochs}
\begin{tabular}{lcccccc}
\hline
Date & Flux density [$\mu$Jy] & S/N & Program ID \\
\hline
25 February 2024 & $115.1\pm 20.8$ & 5.5 & DDTC333 \citep{panda2024lowfrequencywidebandstudyactive} \\
15 June 2024    & $60.5\pm 15.6$ & 3.8 & DDTC351 (this work)  \\
3 July 2024      & $51.1\pm 11.6$ & 4.4 & DDTC353 \citep{panda2024lowfrequencywidebandstudyactive} \\
7 July 2024      & $62.9\pm 19.5$ & 3.2 & DDTC351 (this work)  \\
13 July 2024      & $90.7\pm 24.3$  & 3.7 & DDTC353 \citep{panda2024lowfrequencywidebandstudyactive} \\
28 July 2024     & $74.9\pm 19.5$  & 3.8 & DDTC351 (this work)  \\
22 August 2024   & $70.6\pm 15.4$  & 4.5 & DDTC378 (this work) \\

\hline
\end{tabular}

\begin{flushleft}
\textit{Notes:} Above reported flux densities are peak values measured at the PRS position. The quoted uncertainties represent only the RMS errors around the source position (i.e., only the random errors, with possible contributions from deconvolution systematics), and an additional $\sim$10\% systematic uncertainty (due to the corresponding uncertainty in the flux-density scale used for calibration) is expected. 
\end{flushleft}
\end{table}

\end{document}